\documentclass[preprint,
showpacs, 
amsmath,amssymb, aps,
]{revtex4}

\newcommand{\be}{\begin{equation}}
\newcommand{\ee}{\end{equation}} 
\newcommand{\bea}{\begin{eqnarray}} 
\usepackage{graphicx}
\newcommand{\eea}{\end{eqnarray}}
\usepackage{epsfig}
\usepackage{bm} 

\usepackage{tikz}
\usepackage{dcolumn}
\newcommand{\kbf}{\mathbf{k}}
\newcommand{\vbf}{\mathbf{v}}
\newcommand{\rbf}{\mathbf{r}}

\begin{document}


\title{Nondiagonal Graphene Conductivity in the Presence of In-Plane Magnetic Fields}

\author{R.R. Brand\~ao$^1$ and L. Moriconi$^2$}
\affiliation{$^1$Instituto Nacional da Propriedade Industrial, Rua S\~ao Bento 1, Centro, CEP: 20090-010, Rio de Janeiro, RJ, Brazil}
\affiliation{$^2$Instituto de F\'\i sica, Universidade Federal do Rio de Janeiro,
C.P. 68528, CEP: 21945-970, Rio de Janeiro, RJ, Brazil}


\begin{abstract}
We study the electron/hole transport in puddle-disordered and rough graphene samples which are subject to in-plane magnetic fields. Previous treatments, mostly devoted to regimes where the electron/hole scattering wavelengths are larger than the surface height correlation length, are based on the use of transport equations with appropriate forms for the collision term. We point out in this work, as a counterpoint, that classical Lorentz force effects, which are expected to hold when the Fermi level is far enough away from the charge neutral point, can be heuristically assessed through disordered Boltzmann equations that contain magnetic-field dependent material derivatives, and keep the zero magnetic-field structure of the collision term. It turns out that the electric conductivity tensor gets a peculiar nondiagonal component, induced by the in-plane magnetic field that crosses the rough topography of the graphene sheet, even if the projected random transverse magnetic field vanishes in the mean. Numerical estimates of the transverse conductivities suggest that they are suitable of observation under conditions which are within the reach of up-to-date experimental methods.
\end{abstract}

\maketitle


\section{Introduction}

The high mobility of charge carriers in graphene, related to their pseudo-relativistic Dirac spectrum and semimetal character, render it to be one of the most promising materials for technological innovation in the field of solid state devices \cite{Novoselov-Nature-2012}. A flurry of research has started since the discovery of graphene ten years ago, marked by several important experimental and theoretical advances along the way. There is, however, a spread consensus that the physics of electron transport in graphene is not yet completely understood \cite{DasSarma-RMP-2010}. The simplest of all approaches, viz., a straightforward application of linear response theory to the problem of two-dimensional free massless Dirac fermions is plagued with ambiguities ultimately associated with the evaluation of fermion determinants \cite{Beneventano-J.PhysA-2007}, so that more physical ingredients are in order for proper modeling. A comprehensive study of charge transport in graphene should take into account, as a matter of principle, electron scattering caused by (i) the Coulomb two-body interaction (which seems to affect the fermion spectrum in a relevant way in the vicinity of the charge neutral point), (ii) the disordered substrate doping layer and (iii) surface roughness. At present, one finds in the literature only partial modeling scenarios with variable degrees of success \cite{Caio-Mucciolo-JPCM-2010}.

As it is usual in transport theory, there are essentially two main approaches to the computation of the conductivity tensor. One is based on the Kubo response formalism \cite{Landauer-conductance-PRB-2007}, while the other relies on the analysis of transport equations \cite{DasSarma-RMP-2010}. The latter approach, which is the one to be adopted in this work, is particularly suitable for the investigation of semiclassical regimes. In graphene, these regimes are attained in situations where the Fermi level is far enough from the charge neutral point, or, in equivalent words, for larger absolute charge carrier densities, so that the electron/hole wavelengths involved in scattering can be assumed to be much smaller than the mean free path. 

An interesting transport problem, intimately related to the one of surface roughness characterization, consists in the study of the linear response features of graphene in the presence of in-plane magnetic fields \cite{Lundeberg-Folk-PRL-2010}. As the magnetic field crosses the rough topography of the graphene sample, the electron/hole dynamics couples to the component of the magnetic field that is normal to the surface. Thus, in effective terms, the electron/hole transport can be modeled as if it would take place in the presence of a random transverse magnetic field, with correlations that reflect the random distribution of surface heights in the graphene sheet. If one then restricts the analysis to regimes where scattering wavelengths are taken to be larger than the surface height correlation length, the physical effects of the random transverse magnetic field can all be encoded in the specific form of the collision operator that is introduced in transport equations for the charge carrier density. This is precisely the point of view taken in Refs. \cite{Lundeberg-Folk-PRL-2010,Japoneses-JOP-2011,Genma-Katori-Cont.LundebergFolk-Arxiv-2012, burgos_etal}.

As a counterpoint, our aim in this paper is to explore the magnetic in-plane transport problem in graphene for the case where the scattering wavelengths are smaller than the mean free path and the surface height correlation length, so that we can work within the framework of a semiclassical Boltzmann equation approach, in a spirit similar to what has been done in the context of electron gases \cite{Hedegaard-PRB-1995}. Accordingly, there is no need to modify the structure of the Boltzmann collision term, once the perturbations due to the random transverse magnetic field are given by contributions associated to the classical Lorentz force, which appear in the left hand side of the Boltzmann equation. The central result of our analysis is the peculiar nondiagonal structure of the conductivity tensor, which is closely connected to the statistical properties of the graphene rough surface.

This paper is organized as follows. To start, a gaussian model of surface roughness is introduced in Sec. II, which is then used to establish the statistical properties of the effective random transverse magnetic field defined on the graphene sheet. In Sec. III, we develop, following the Boltzmann equation approach of Ref. \cite{LucaDaniel-PRB-2011}, a treatment of the semiclassical electron/hole transport in graphene samples disordered by the presence of charged puddles \cite{DasSarma} and subject to in-plane magnetic fields. We, then, work out numerical estimates of the conductivity tensor components, which are noted to be perfectly within the reach of present experimental resolution. Finally, in Sec. IV, we summarize our findings and point out directions of further research.

\section{Random Gaussian Model of Surface Roughness}

We are interested to study charge transport in a rough graphene sheet which is subject to crossed in-plane electric and magnetic fields. Surface heights can be represented as a real function $z=z(\rbf)$, where $\rbf = (x,y)$ denotes a point defined on the $xy$ plane. We take, without loss of generality, the $x$ direction to be parallel to the external (in-plane) small electric field $\bold E$ that is used to probe the charge transport in the sample, that is, $\bold E \equiv (E,0,0)$. For modeling purposes, $z(\bold r)$ is assumed to be a smooth gaussian random field with vanishing expectation value, $\langle z(\bold r) \rangle =0$, and two-point correlation function
\begin{equation}
\langle z\left(\bold r \right)z\left(\bold r' \right)\rangle = \left\{
\begin{array}{l}
\frac{H^2}{2} \left ( 1 - \frac{2|\bold r - \bold r'|^2}{L^2}
+\frac{|\bold r - \bold r'|^4}{L^4} \right ) \ , \  {\hbox{ if }} |\bold r - \bold r'| \leq L \ , \ \\
0 \ , \ {\hbox{ if }} |\bold r - \bold r'| > L \ , \
\end{array}
\right.
\end{equation}
where $H$ and $L$ parametrize the standard deviation and the correlation length of height fluctuations on the graphene sheet, respectively. The unit normal vector on the graphene surface at position $\bold r$ can be written as
\begin{equation}
\hat n(\bold r) = \frac{1}{\sqrt{1+a^2+b^2}} (-a,-b,1) \simeq (-a,-b,1) \ , \ \label{n_vec}
\end{equation}
where $a \equiv \partial_x z(\bold r)$, $b \equiv \partial_y z(\bold r)$ and the approximation in (\ref{n_vec}) is related to the surface smoothness assumption ($a, b \ll 1$).

Consider now, that the in-plane magnetic field $\bold B$ applied on the graphene sample makes an angle $\phi$ with the electric field. We have, thus,
\begin{equation}
\bold B = B_0 (\cos(\phi), \sin(\phi),0) \ . \ \label{inplaneB}
\end{equation}
The component of the magnetic field that is parallel to the graphene surface has not any role in the dynamics of charge transport. The normal component, on the other hand, 
\begin{equation}
B(\bold r) \equiv \bold B \cdot \hat n(\bold r) = -B_0 \left(\cos\left(\phi\right) \partial_x z + \sin\left(\phi\right) \partial_y z \right) \ , \ \label{randomB}
\end{equation}
is a random field that couples to the orbital electron/hole degrees of freedom. Since $z=z(\bold r)$ is a random gaussian field, so it is the transverse magnetic $B=B(\bold r)$, which vanishes in the mean, i.e., $\langle B(\rbf) \rangle =0$. We can compute, from Eq. (\ref{randomB}), arbitrary N-point expectation values like
\begin{equation}
G_N(\bold r_1, \bold r_2,...,\bold r_N) \equiv \langle B(\bold r_1)B(\bold r_2) ... B(\bold r_N) \rangle \ . \
\end{equation}
It is clear that $G_N=0$ for $N$ odd. We list, below, expectation values that are of particular importance in our analysis:
\begin{eqnarray}
&&\langle (B(\bold r))^2 \rangle = G_2(\bold r, \bold r) = \frac{2H^2}{L^{2}}B_{0}^{2}\ ,\label{b} \ \\
&&\langle\left(\partial_x B(\bold r) \right )^2 \rangle  = 
\lim_{\bold r' \rightarrow \bold r} \partial_x \partial_{x'} G_2(\bold r, \bold r') =
\frac{4H^{2}}{L^{4}}B_{0}^{2}\left(\sin^{2}\left(\phi\right)+3\cos^{2}\left(\phi\right)\right) \ ,\label{dxb} \\\
&&\langle\left(\partial_y B(\bold r) \right )^2 \rangle  = 
\lim_{\bold r' \rightarrow \bold r} \partial_y \partial_{y'} G_2(\bold r, \bold r') =
\frac{4H^{2}}{L^{4}}B_{0}^{2}\left(\cos^{2}\left(\phi\right)+3\sin^{2}\left(\phi\right)\right) \ , \label{dyb}\ \\
&&\langle \partial_x B(\bold r) \partial_y B(\bold r) \rangle  = 
\lim_{\bold r' \rightarrow \bold r} \partial_x \partial_{y'} G_2(\bold r, \bold r') =
\frac{4H^{2}}{L^{4}}B_{0}^{2}\sin\left(2 \phi\right)
 \ . \
\label{dxdyb}
\end{eqnarray}
The effective random magnetic field $B(\rbf)$ is not statistically isotropic, that is $\langle B(\rbf) B(\rbf') \rangle$ is not a function of $|\rbf - \rbf'|$. In fact, we have, for $|\rbf| < L$,
\begin{equation}
 \langle B(0) B(\rbf) \rangle = 2 \frac{B_0^2 H^2}{L^2} \left \{ 1 - \frac{2}{L^2} \left [r^2+2( x \cos (\phi) + y \sin(\phi) )^2 \right ] \right \}
  \ . \ \label{BB}
\end{equation}
As derived in the next section, the anisotropic correlation function (\ref{BB}) leads, ultimately, to a non-diagonal conductivity tensor that depends on the geometrical parameters $\phi$, $H$, and $L$.

It is instructive to briefly digress on the apparently analogous unusual Hall effect, which is predicted to occur, under very special circumstances, in the completely different context of topological insulators \cite{Haldane-Original, Haldane-Wright}, in a model where the magnetic field also vanishes in the mean. As it is well-known, the existence of Hall response is necessarily associated with time-reversal symmetry breaking (in two-dimensional space), which in the aforementioned model is broken due to a specific distribution of magnetic fluxes around lattice links. In our case, in contrast, the essential point in having non-diagonal conductivities is related to a peculiar parity-symmetry breaking mechanism. Actually, one could be puzzled by the fact that the magnetic field vanishes in the mean -- as it would be implied by parity-reversal symmetry. Therefore, the transverse conductivity, which is odd under parity-reversal transformations, should vanish as well. However, it is important to note that parity-symmetry is broken in our setting only at the level of second-order correlation functions. In fact, parity-symmetry reversal can be implemented in our analysis very simply by means of the replacement  
$\phi \rightarrow - \phi$ in (\ref{inplaneB}), modifying, as a consequence, the expectation values (\ref{dxdyb}) and (\ref{BB}).

\section{Semiclassical Transport in Disordered Graphene}

In the absence of external magnetic fields, the usual (one-body) modelling ingredients in the graphene charge transport problem are the electron/hole scattering by substrate impurities and the smooth random electric potential associated to charged puddles. In the Boltzmann equation approach, one assumes that scattering by impurities is encoded in the relaxation time approximation, while charged puddles can be modeled as extended subregions of the sample where the chemical potential is approximately uniform, but randomly fluctuating from puddle to puddle. In that way, a phenomenological relation between the minimum conductivity value, the electron/hole mobility parameter (in the semiclassical region) and the steepness of the conductivity parabola around the charge neutral point has been predicted and clearly supported by an extensive compilation of experimental data \cite{LucaDaniel-PRB-2011}.

We extend in this work the Boltzmann equation approach put forward in \cite{LucaDaniel-PRB-2011} to the more general context of graphene transport in the presence of in-plane magnetic fields, where, as discussed in the preceding section, surface roughness becomes an additional source of disorder. We deal, more specifically, with approximately semiclassical transport regimes characterized by a mean free path $\ell_k$ and a surface correlation height $L$ which are both larger than the scattering wavelength $\sim k^{-1}$ and both smaller than the typical charged puddle linear size $L_p$, that is,
\begin{equation}
k^{-1} < \ell_k < L_p \ , \ k^{-1} < L < L_p  \ . \ \label{scales}
\end{equation}
The partition of the sample into self-correlated magnetic field domains and charged puddles is depicted in Fig. 1.

Of course, the Boltzmann equation approach would be firmly grounded if $k^{-1} \ll \ell_k$. However, as we will see, usual graphene samples and properly chosen carrier concentrations provide us at most with a ``weak" separation of scales at the borderline of semiclassical behavior. Our discussion, thus, is essentially heuristic, having in mind the present lack of understanding on the crossover region between the semiclassical 
and the quantum regimes of graphene transport.

We define, accordingly to the first set of inequalities in (\ref{scales}), a stationary Boltzmann distribution function $f^\pm (\mathbf{k} , \mathbf{r},\xi)$ in each one of the charged puddles, where $\kbf$ and $\rbf$ are, respectively, the wavenumber and position vectors, $\xi$ denotes the ``puddle-dependent" shift of the chemical potential, and the positive and negative superscripts refer, respectively, to holes and electrons. Working in the relaxation time approximation, we write, for a given puddle of area $A$, the Boltzmann transport equation as
\begin{equation}
\left \{ \pm \frac{e}{\hbar}\left[\mathbf{E}+ B(\rbf) \mathbf{v_{k}}\times \hat z  \right]
\cdot \nabla_{\mathbf{k}}
+ \mathbf{v_{k}} \cdot \nabla_{\mathbf{r}} \right \} f^\pm (\mathbf{k} , \mathbf{r},\xi)
= - \frac{1}{\tau_{\mathbf{k}}} [f^\pm (\mathbf{k}, \mathbf{r},\xi)-f^\pm_0 (\mathbf{k} , \xi)] \ , \ 
\label{subst}
\end{equation}
where
\begin{equation}
f^\pm_0 (\mathbf{k} , \xi) \equiv 4A^{-1} \Theta(\xi \pm \mu - \epsilon_k)
\end{equation}
is the zero-temperature Fermi-Dirac distribution for an ideal gas of holes or electrons with chemical potential $\xi \pm \mu$, which also takes into account the valley and the spin degrees of freedom (Zeeman splitting effects are negligible in our study). The one-particle energy spectrum is given by $\epsilon_{k}=ak^{\alpha}$ (where $\alpha=1$ and $\alpha=2$ are assumed to model monolayer and bilayer graphene systems, respectively). The wavenumber dependent particle/hole velocity is, therefore,
$\vbf_\kbf = \hbar^{-1} \nabla_\kbf \epsilon_\kbf = \alpha a \hbar^{-1} k^{\alpha - 2} \kbf$. To evaluate $a$ in the particular case of monolayer graphene, we recall that the observed Fermi velocity is $v_F = a/ \hbar \simeq 10^6 $m/s. 

Still following Ref. \cite{LucaDaniel-PRB-2011},
the relaxation time is defined as
\begin{equation}
\tau_{k}=\frac{c \hbar k^{2-\alpha}}{\alpha an_{imp}} \ , \ \label{tau}
\end{equation}
where $n_{imp}$ is the concentration of scattering impurities and $c \simeq 1.6$ is a dimensionless prefactor (the underlying model is the one of coulombian impurities spread over a $SiO_{2}$ substrate; $c$ is a function of Wigner-Seitz radius \cite{DasSarma}). It is known that (\ref{tau}) leads, in the absence of external magnetic field, to conductivity profiles that depend linearly on the charge carrier density in the semiclassical regime, as corroborated in real experiments.
\vspace{-0.5cm}

\begin{center}
\begin{figure}
\includegraphics[scale=0.5]{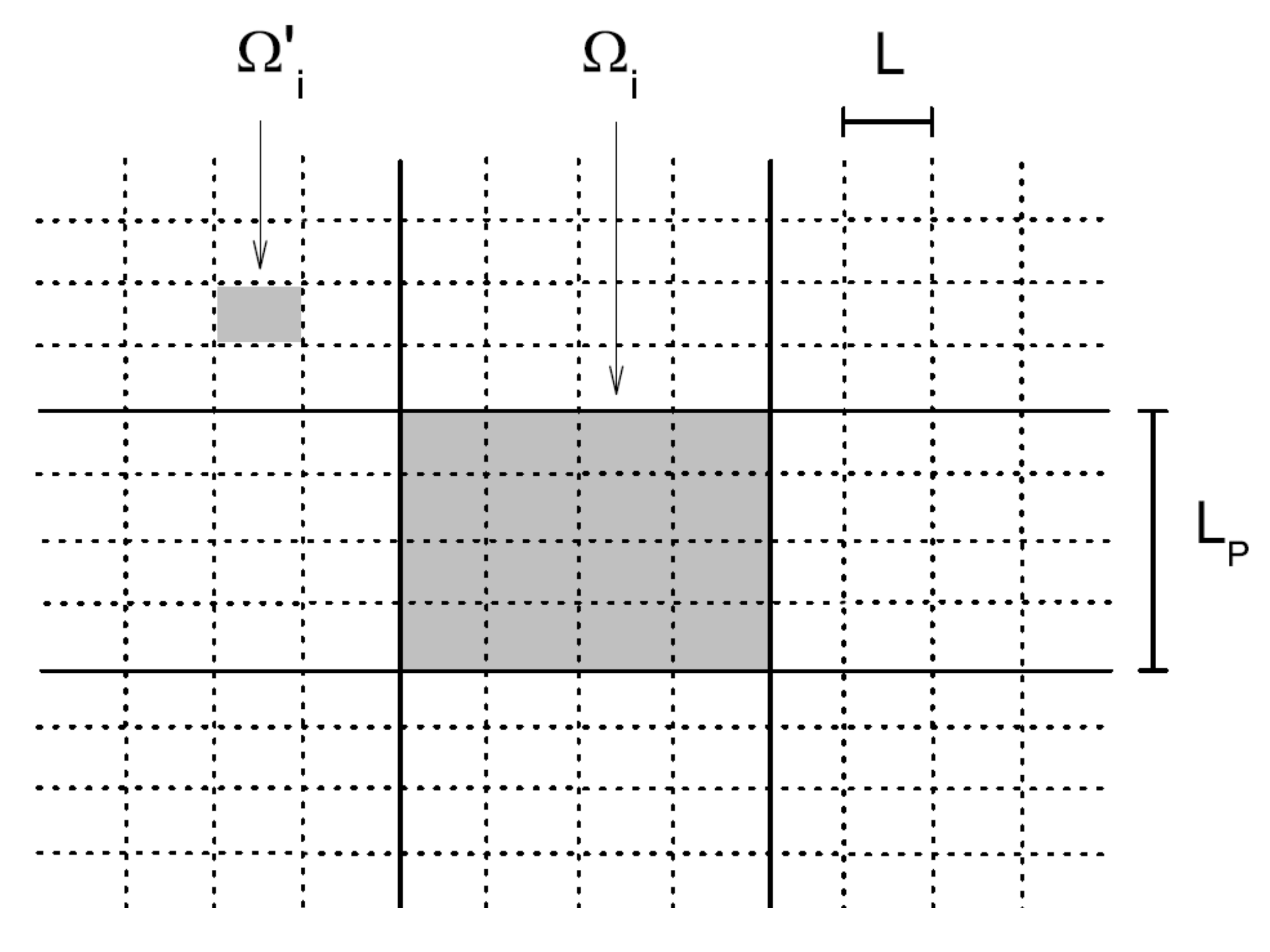}
\caption{Conceptual scheme of the graphene model for the analysis of disordered semiclassical transport. The sample is partitioned into charged puddle domains $\Omega_i$ (bounded by solid lines) and spatially smooth subdomains $\Omega'_i$ (bounded by dotted lines) of typical linear sizes $L_P$ and $L$, respectively.}
\label{model}
\end{figure}
\end{center}

The random chemical potential shift $\xi$ is assumed to have zero-mean and standard deviation $\delta \epsilon_0$, which introduces an energy scale into the problem, related to displacements of the energy band across the graphene sample due to the existence of charged puddles. It has been put forward in Ref. \cite{LucaDaniel-PRB-2011}, with reasonable phenomenological success, that the probability distribution function (pdf) of $\xi$ can be written as 
\begin{equation}
\rho(\xi) = \frac{1}{\delta \epsilon_0} g \left (\frac{\xi}{\delta \epsilon_0} \right ) \ , \ 
\label{pdf}
\end{equation}
where $g(\cdot)$ is a universal (i.e., sample independent) pdf with unit variance (an educated guess is to
take it as a gaussian).

Once Eq. (\ref{subst}) is solved, one may compute the electric conductivity tensor components as
\begin{eqnarray}
&&\sigma_{xx} = \sigma_{yy} = \frac{e}{2 \pi^2}  \frac{\partial}{\partial E} \left \{ 
\int d^2 \kbf d^2 \rbf \langle \overline{[f^+ (\kbf , \rbf, \xi) - f^- (\kbf , \rbf, \xi)]}
\rangle \hat x \cdot \vbf_\kbf \right \}_{E=0} \ , \ \nonumber \\
&&\sigma_{xy} = - \sigma_{yx} = \frac{e}{2 \pi^2}  \frac{\partial}{\partial E} \left \{ 
\int d^2 \kbf d^2 \rbf \langle \overline{[f^- (\kbf , \rbf, \xi) - f^+ (\kbf , \rbf, \xi)]}
\rangle \hat y \cdot \vbf_\kbf \right \}_{E=0} \ , \ \nonumber \\
\label{conduct_xx_xy}
\end{eqnarray}
where the above double-average notation stands for the computation of expectation values in the ensemble of external random magnetic fields $B(\rbf)$ (expression between brackets) and random chemical potential shifts $\xi$ (expression embraced by an overbar). It has been tacitly assumed, from the definition (\ref{pdf}), that fluctuations of the chemical potential (and therefore charged puddles) are not correlated to corrugations of the graphene sheet. This is an interesting issue, still open to theoretical and experimental discussion. Actually, it has been suggested that charged puddles could be found as the sole effect of graphene sheet roughness, even if they are not coupled to substrates as in suspended graphene samples \cite{gilbertini_etal}. In this work, we take the much simpler phenomenological view encoded in (\ref{pdf}). Otherwise, we should work with conjectured joint probability distribution functions defined on the sample space of chemical potential and surface height fluctuations. As a first approximation, however, it turns out that chemical potential fluctuations give subleading corrections to the conductivity tensor, so in this first approach to the problem we do not have worry too much with the precise modeling of the joint random fluctuations of $\xi$ and $z(\rbf)$.

It is possible, then, to  perform the average over fluctuations of $\xi$ straightforwardly in (\ref{subst}), which means that we just replace, in that equation, $f^\pm (\kbf, \rbf, \xi)$ and $f^\pm_0( \kbf , \xi)$ by, respectively,
\begin{equation}
f^\pm (\kbf , \rbf) \equiv \overline{f^\pm (\kbf , \rbf, \xi)} =  \int_{- \infty}^{\infty} d \xi \rho(\xi) f^\pm (\kbf , \rbf, \xi)
\end{equation}
and
\begin{equation}
f^\pm_0(\kbf) \equiv  \overline{f^\pm_0 (\kbf , \rbf, \xi)}
= \int_{- \infty}^{\infty} d \xi \rho( \xi) f^\pm_0 (\kbf , \xi) \ . \ \label{f0avg}
\end{equation}
The Boltzmann equation can be formally solved as
\begin{equation}
f^\pm (\mathbf{k} , \rbf) = [ 1+D^\pm (\kbf , \rbf) ]^{-1}f^\pm_0 (\mathbf{k}) = 
\left \{ 1+ \sum_{n=1}^\infty [-D^\pm (\kbf , \rbf)]^n \right \} f^\pm_0 (\mathbf{k}) \ , \   \label{fkk0}
\end{equation}
where 
\begin{equation}
D^\pm (\kbf , \rbf) = \pm \frac{e \tau_\kbf }{\hbar}\left[\mathbf{E}+ B(\rbf) \mathbf{v_{k}}\times \hat z  \right]
\cdot \nabla_{\mathbf{k}}
+ \tau_\kbf \mathbf{v_{k}} \cdot \nabla_{\mathbf{r}} \ . \
\label{D+-}
\end{equation}
Relying upon the scaling forms of (\ref{tau}) and of the energy spectrum, it is not difficult to show that
\begin{eqnarray}
&&\int d^2 \kbf  \langle [D^\pm(\kbf, \rbf)]^n \rangle f^\pm_0(\kbf) \nonumber \\
&& =   E \sum_{p=0}^{\frac{n-1}{2}} 
C^\pm(n,p) \int d^2 \kbf \frac{e \tau_\kbf}{\hbar k}  \left ( \frac{\ell_k}{L} \sqrt{\frac{H}{\ell_c}} \right )^{n-1}
\left (\frac{H}{\ell_c} \right )^p f^\pm_0(\kbf) + {\cal{O}}(E^2) \ , \
\label{approx}
\end{eqnarray}
where $C^\pm(n,p)$ is dimensionless, $\ell_c = \hbar k/ eB_0$ and $\ell_k = |\vbf_\kbf| \tau_\kbf$ are, respectively, the rms cyclotron radius and the mean free path, both referring to excitations with wavenumber $k$. In usual graphene samples, we have typically $H < 0.5$ nm and $L > 8$ nm. A proper choice of the Fermi wavenumber in the semiclassical region leads to $\ell_c > 20$ nm, for $B_0 < 5$ T, and $\ell_k < 30$ nm in the integrand of (\ref{approx}). It turns out that
\begin{equation}
\frac{\ell_k}{L} \sqrt{{\frac{H}{\ell_c}}} < 0.6
\end{equation}
and
\begin{equation}
\frac{H}{\ell_c} < 2.5 \times 10^{-2}
\end{equation}
are in fact small enough to suggest the (asymptotic) convergence of the perturbative expansion (\ref{approx}), 
which is then carried up to order $n=5$, with $p=0$, so that we are able to find the leading anisotropic contributions to the conductivity. 
We get
\begin{eqnarray}
\sigma_{xx} =
\frac{A}{4 \pi} 
\int_0^\infty dk &&\left\{ - \frac{e^{2} c k^2}{\hbar n_{imp}}
+\frac{e^{4}c^{5}k^4}{\hbar^{3} n_{imp}^{5}}
\left \langle 2[\partial_x B(\rbf) ] ^{2}+[ \partial_y B (\rbf) ]^{2} \right \rangle \right. \nonumber \\
&& \left. +\frac{e^{4} c^{3}k^2}{ \hbar^{3}n_{imp}^{3}} \left \langle [B(\rbf)]^2
\right \rangle \right \} \frac{d}{dk} [f^+_0(\kbf) - f^-_0(\kbf)]
\label{sigmaxx}
\end{eqnarray}

and
\begin{equation}
\sigma_{xy} =
\frac{Ae^{4}c^{5}}{4 \pi \hbar^{3} n_{imp}^{5}} \left \langle \partial_x B(\rbf) \partial_y B(\rbf)  \right \rangle 
\int_0^\infty dk k^4
\frac{d}{dk} [f^-_0(\kbf) - f^+_0(\kbf)] \ . \
\label{sigmaxy}
\end{equation}

Substituting, now, Eqs. (\ref{b}) to (\ref{dxdyb}) and (\ref{f0avg}) in (\ref{sigmaxx}) and (\ref{sigmaxy}), it follows that
\begin{eqnarray}
\sigma_{xx}&=&\frac{c e^{2}}{2\pi L^{4}\hbar^{3}n_{\text{imp}}^{5}}\int_{- \infty}^{\infty}d\xi\rho(\xi) \left [L^{2}n_{\text{imp}}^{2}\left(\frac{\mu+\xi}{a}\right)^{2/\alpha}\left(L^{2}\hbar^{2}n_{\text{imp}}^{2}-2B_{0}^{2}H^{2}c^{2}e^{2}\right) \right.
\nonumber \\
&+&\left(\frac{\xi-\mu}{a}\right)^{2/\alpha}\left(2B_{0}^{2}H^{2}c^{2}e^{2}\left(c^{2}\left(\cos\left(2\phi\right)+10\right)\left(\frac{\xi-\mu}{a}\right)^{2/\alpha}-L^{2}n_{\text{imp}}^{2}\right)+L^{4}\hbar^{2}n_{\text{imp}}^{4}\right) \nonumber \\
 &+& \left. \left(\frac{\mu+\xi}{a}\right)^{4/\alpha}2B_{0}^{2}H^{2}c^{4}e^{2}\left(\cos\left(2\phi\right)+10 \right) \right]
 \label{sigmaxxfinal}
\end{eqnarray}
and
\begin{equation}
\sigma_{xy}= \frac{4B_{0}^{2}H^{2}e^{4}c^{5} \sin(2\phi)}{\pi L^{4}\hbar^{3}n_{\text{imp}}^{5}} \int_{-\infty}^\infty d\xi  \rho(\xi)
\left[ \left (\frac{\xi-\mu}{a} \right)^{4/\alpha}+\left(\frac{\xi+\mu}{a} \right)^{4/\alpha} \right ] 
\ . \ 
 \label{sigmaxyfinal}
\end{equation}
Eq. (\ref{sigmaxyfinal}) is odd under parity-reversal transformations, as it should be. In fact, following the discussion of Sec. II, parity-reversal is implemented in (\ref{sigmaxyfinal}) by means of the substitution $\phi \rightarrow - \phi$, which changes the sign of $\sigma_{xy}$. The evaluation of (\ref{sigmaxxfinal}) and (\ref{sigmaxyfinal}) is performed with the help of (\ref{pdf}) and two useful relations taken from Ref. \cite{LucaDaniel-PRB-2011}, viz.,
\begin{equation}
n_{0}=\frac{1}{2\pi}\left(\frac{\delta\epsilon_{0}}{a}\right)^{\frac{2}{\alpha}}\left(\int\xi^{\frac{2}{\alpha}}g(\xi)\, d\xi\right)
\end{equation}
and
\begin{equation}
\frac{\mu^2}{a^2} = 2 \pi |n - n_0| \ , \
\end{equation}
where $n$ and $n_0$ are the carrier densities defined at chemical potential $\mu$ and at the charge neutral point ($\mu=0$). We note that although $\delta \epsilon_0$ it is not known a priori, we would need to know its value only in eventual contributions to (\ref{sigmaxxfinal}) and (\ref{sigmaxyfinal}) which are of the order of $(\delta \epsilon_0 / \mu)^4$, and, therefore, are assumed to be negligible in the semiclassical regime.

For a proper physical interpretation of the results (\ref{sigmaxxfinal}) and (\ref{sigmaxyfinal}), we have to keep in mind that the conductivity corrections found here for the semiclassical regime are dominated by effects related to the random topography of the graphene sheet. Charged puddle effects, on the other hand, are more relevant in situations where the chemical potential is close enough to the charge-neutral point \cite{LucaDaniel-PRB-2011}. Thus, if we completely neglect fluctuations of the chemical potential, taking the limit $\delta \epsilon_0 \rightarrow 0$ in (\ref{sigmaxyfinal}), using (\ref{pdf}), it is not difficult to show that the transverse conductivity (\ref{sigmaxyfinal}) is still non-vanishing and given by
\begin{equation}
\sigma_{xy}= \frac{8B_{0}^{2}H^{2}e^{4}c^{5} \sin(2\phi)}{\pi L^{4}\hbar^{3}n_{\text{imp}}^{5}}
\left (\frac{\mu}{a} \right)^{4/\alpha}
\ . \ 
\end{equation}

\begin{center}
\begin{figure}[h!]
\includegraphics[scale=0.58]{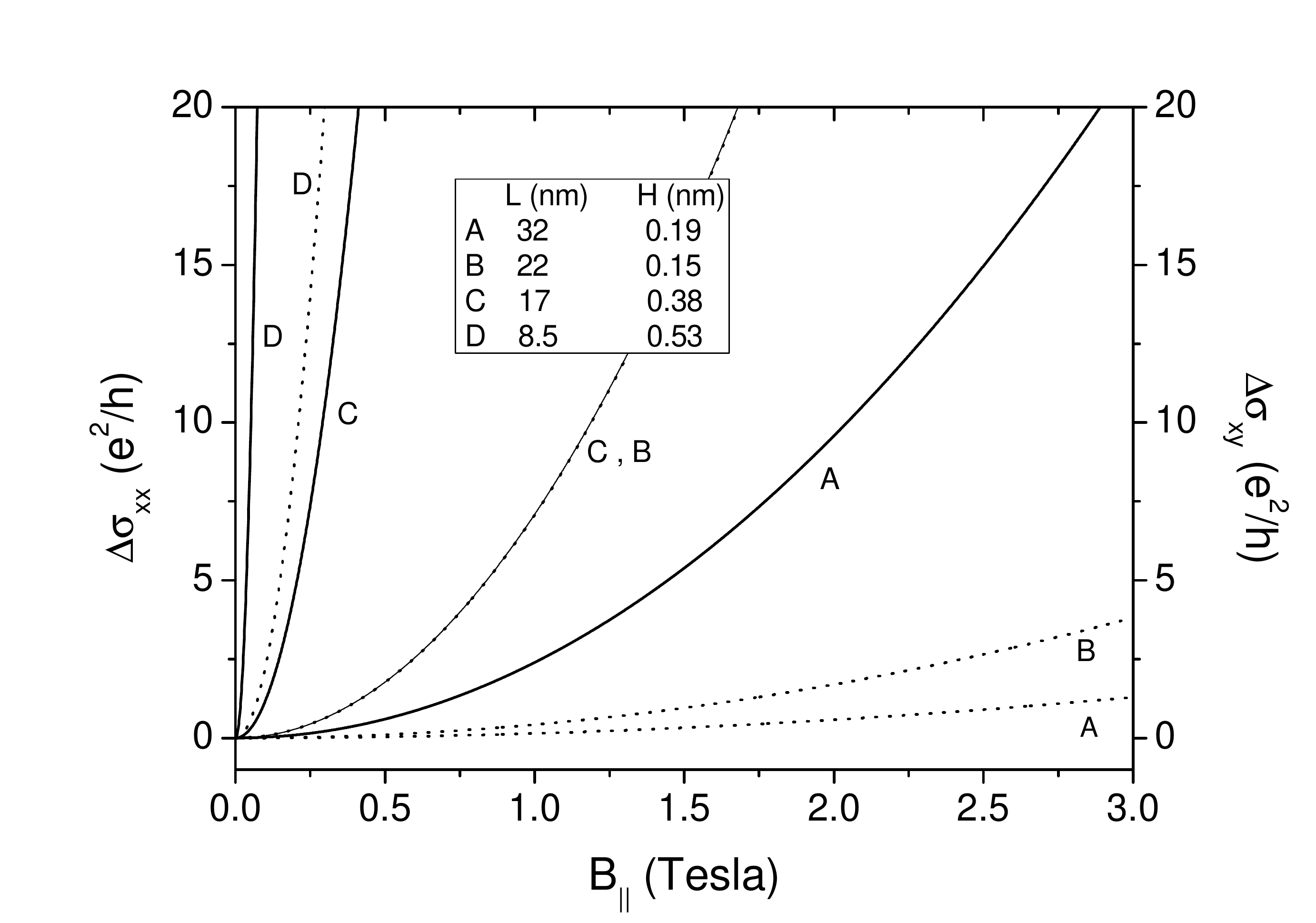}
\caption{Longitudinal (solid lines) and transverse (dotted lines) conductivity profiles, for $n_{imp} = 2 \times 10^{11}$ cm$^{-2}$ and carrier density $n = 5 \times 10^{11}$ cm$^{-2}$, evaluated with input data $L$ and $H$ taken from Refs \cite{Ishigami-NanoLetters-2007} (A), \cite{luiultraflat2009} (B), and \cite{Genma-Katori-Cont.LundebergFolk-Arxiv-2012} (C and D).}
\label{model2}
\end{figure}
\end{center}

We report here results for the particular case of monolayer graphene (there are no further technical difficulties in dealing with bilayer graphene). We plot, in Fig. \ref{model2},  the corrections $\Delta \sigma_{xx}$ and $\Delta \sigma_{xy} \equiv \sigma_{xy}$ to the longitudinal and transverse conductivities, respectively, induced by the in-plane magnetic field, using surface roughness data from the current literature. The angle between the magnetic and the electric field is taken to be $\phi = \pi/4$, in order to maximize the transverse conductivity, as given in (\ref{sigmaxyfinal}). In our numerical estimates, we have made use of the phenomenological expression $n_0 \simeq n_{imp}/10$, which relates the impurity and carrier concentrations at the charge neutral point \cite{zhangorigin2009}. In all of the profiles, the longitudinal conductivity in the absence of the in-plane magnetic field is estimated as $\sigma_{xx} \simeq 100 e^2 /h$ (observe that graphene conductivities as larger as $300 e^2 /h$ have already been addressed in the literature \cite{tan_etal, anicic_miskovic}). Also, it is likely that the inequalities (\ref{scales}) can be experimentally realized, since in our case study (devised for carrier density $n = 5 \times 10^{11}$ cm$^{-2}$), we have $k^{-1} \simeq$ 5 nm , 20 nm $< \ell_k <$ 30 nm, 10 nm $< L <$ 30 nm, and, from a number of scanning tunneling experiments, 30 nm $< L_p <$ 50 nm. \cite{yacoby_etal,crommie_etal,leroy_etal}.

It is clear that the peculiar non-vanishing transverse conductivity $\sigma_{xy}$ predicted and estimated here, the central result of this paper, can be resolved by the present conductivity measurement techniques, which are able to record fractions of $e^2/h$ \cite{Lundeberg-Folk-PRL-2010,Japoneses-JOP-2011,Genma-Katori-Cont.LundebergFolk-Arxiv-2012}. We also stress that the range of magnetic field intensities investigated in Fig. 2 is even smaller than the ones currently used in graphene transport experiments \cite{Lundeberg-Folk-PRL-2010,Japoneses-JOP-2011,Genma-Katori-Cont.LundebergFolk-Arxiv-2012}.

\section{Conclusions}

We have studied, within the semiclassical Boltzmann equation approach, charge transport in usual
graphene samples which have both charged puddle domains (caused by the interaction with a substrate) 
and rough surface profiles. Our particular interest is related to situations where an in-plane magnetic field
is applied to the sample, in such a way that charge transport takes place, effectively, under the presence of a random
transverse magnetic field. The key point in our analysis is to consider regimes where scattering wavelenghts are smaller 
than the surface height correlation length, so that all the magnetic field effects can be brought, as Lorentz force contributions, 
to the left hand side of the Boltzmann transport equation. A straightforward perturbative expansion, leads, then, to
the corrected conductivity tensor. The somewhat surprising result, which to the authors' knowledge has not yet
been explored in the graphene literature, is the prediction of a non-vanishing transverse conductivity,
without mean external magnetic field. This phenomenon, which is in principle within 
the reach of current experimental techniques, is likely to be relevant in studies of graphene surface characterization, 
once the conductivity tensor turns out to depend on combinations of the surface statistical parameters $H$ and $L$.

It would be interesting, as a topic for further research, to address the possible existence of similar magnetic-induced 
anisotropic effects in fully quantum regimes where the in-plane magnetic field is assumed to affect only the right hand side 
of the Boltzmann equation. Also, one may wonder how the transition from the semiclassical to the quantum regime should be 
modeled with the help of transport equations, which is, no doubt, a challenging theoretical problem.

\acknowledgments

We are greatly indebted to Caio Lewenkopf for several enlightening discussions and Eduardo Marino for calling our attention
to Ref. \cite{Haldane-Original}. This work has been partially supported by CNPq and FAPERJ.


\begin{references}

\bibitem{Novoselov-Nature-2012} K. Novoselov, V. Falko, L. Colombo, P. Gellert, M. Schwab, and K. Kim, Nature {\bf{490}}, 192 (2012).

\bibitem{DasSarma-RMP-2010} S. Das Sarma, S. Adam, E. H. Hwang, and E. Rossi,  Rev. Mod. Phys. {\bf{83}}, 407 (2011).

\bibitem{Beneventano-J.PhysA-2007} C.G. Beneventano, P. Giacconi, E.M. Santangelo, and R. Soldati, J. Phys. {\bf{A40}}, F435 (2007).

\bibitem{Caio-Mucciolo-JPCM-2010} E.R. Mucciolo and C.H. Lewenkopf,  J. Phys. Cond. Matter {\bf{22}}, A263201 (2010).

\bibitem{Landauer-conductance-PRB-2007} S. Ryu, C. Mudry, A. Furusaki, and A.W.W. Ludwig, Phys. Rev. B {\bf{75}}, 205344 (2007).

\bibitem{Lundeberg-Folk-PRL-2010} M.B. Lundeberg and J.A. Folk, Phys. Rev. Lett. {\bf{105}}, 146804 (2010).

\bibitem{Japoneses-JOP-2011} J. Wakabayashi and T. Sano, J. Phys.: Conf. Series {\bf{334}}, 012039 (2010).

\bibitem{Genma-Katori-Cont.LundebergFolk-Arxiv-2012} K. Genma and M. Katori, http://arxiv.org/abs/1211.2046.

\bibitem{burgos_etal} R. Burgos, J. Warnes, L.R.F. Lima, and C. Lewenkopf, Phys. Rev. B {\bf{91}}, 115403 (2015).

\bibitem{Hedegaard-PRB-1995} P. Hedegaard and A. Smith, Phys. Rev. B {\bf{51}}, 10869 (1995).

\bibitem{LucaDaniel-PRB-2011} L. Moriconi and D. Niemeyer, Phys. Rev. B {\bf{84}}, 193401 (2011).

\bibitem{DasSarma} S. Adam, E.H. Hwang, V.M. Galitski, and S. Das Sarma, Proc. Nat. Acad. Sci. {\bf{104}}, 18392 (2007).

\bibitem{Haldane-Original} F.D.M. Haldane, Phys. Rev. Lett. {\bf{61}}, 2015 (1988).

\bibitem{Haldane-Wright} A.R. Wright,  Sci. Rep. {\bf{3}}, 2736 (2013).

\bibitem{gilbertini_etal} M. Gibertini, A. Tomadin, F. Guinea, M.I. Katsnelson, and M. Polini, Phys. Rev. B {\bf{85}}, 201405(R) (2012).

\bibitem{zhangorigin2009} Y. Zhang, V. W. Brar, C. Girit, A. Zettl, and M.F. Crommie, Nat. Phys. {\bf{5}}, 722 (2009).

\bibitem{Ishigami-NanoLetters-2007} M. Ishigami, J.H. Chen, W.G. Cullen, MS. Fuhrer, and E.D. Williams, Nano Lett. {\bf{7}}, 1643 (2007).

\bibitem{luiultraflat2009} C.H. Lui, L. Liu, K.F. Mak, G.W. Flynn, and T.F. Heinz, Nature {\bf{462}}, 339 (2009). 

\bibitem{tan_etal} Y.-W. Tan, Y. Zhang, K. Bolotin, Y. Zhao, S. Adam, E.H. Hwang, S. Das Sarma, H.L. Stormer, and 
P. Kim, Phys. Rev. Lett. {\bf{99}}, 246803 (2007).

\bibitem{anicic_miskovic} R. Anicic and Z.L. Miskovic, Phys. Rev. B {\bf{88}}, 205412 (2013).

\bibitem{yacoby_etal} J. Martin, N. Akerman, G. Ulbricht, T. Lohmann, J.H. Smet, K. von Klitzing, and A. Yacoby, Nature Phys., {\bf{4}}, 144 (2008).

\bibitem{crommie_etal} Y. Zhang, V.W. Brar, C. Girit, A. Zettl, and M.F. Crommie, Nature Phys. {\bf{5}}, 722 (2009).

\bibitem{leroy_etal} A. Deshpande, W. Bao, Z. Zhao, C. N. Lau, and B. J. LeRoy, Phys. Rev. B {\bf{83}}, 155409 (2011).

\end{references}
\end{document}